\documentclass[preprintnumbers,amsmath,amssymb,floatfix,12pt,superscriptaddress,nofootinbib]{revtex4}
\usepackage{graphicx}
\usepackage{epsfig}
\usepackage{bm}
\usepackage{amsfonts}

\begin{document}

\title{FRW Cosmology with Variable $G$ and $\Lambda$}

\author{Mubasher Jamil}
\email{mjamil@camp.nust.edu.pk} \affiliation{Center for Advanced
Mathematics and Physics, National University of Sciences and
Technology, H-12, Islamabad, Pakistan.}

\author{Ujjal Debnath}
\email{ujjaldebnath@yahoo.com , ujjal@iucaa.ernet.in}

\affiliation{Department of Mathematics, Bengal Engineering and
Science University, Shibpur, Howrah-711 103, India.\\}

\begin{abstract}
\vspace*{1.5cm} \centerline{\bf Abstract} \vspace*{1cm} We have
considered a cosmological model of the FRW universe with variable
$G$ and $\Lambda$. The solutions have been obtained for flat model
with particular form of cosmological constant. The cosmological
parameters have also been obtained for dust, radiation and stiff
matter. The statefinder parameters are analyzed and have shown
that these depends only on $w$ and $\epsilon$. Further the
lookback time, proper distance, luminosity distance and angular
diameter distance have also been calculated for our model.
\\ \\
\textbf{Keywords:} Newton's gravitational constant; cosmological
constant; cosmology.
\end{abstract}

 \maketitle

\newpage

\section{Introduction}

The Einstein field equation has two parameters - the gravitational
constant $G$ and the cosmological constant $\Lambda$. The Newtonian
constant of gravitation $G$ plays the role of a coupling constant
between geometry and matter in the Einstein field equations. In an
evolving Universe, it appears natural to look at this ``constant''
as a function of time. Numerous suggestions based on different
arguments have been proposed in the past few decades in which $G$
varies with time \cite{wess,tiwari}. Dirac \cite{Dir} proposed a
theory with variable $G$ motivated by the occurrence of large
numbers discovered by Weyl, Eddington and Dirac himself. Many other
extensions of Einstein's theory with time dependent $G$ have also
been proposed in order to achieve a possible unification of
gravitation and elementary particle physics or to incorporate Mach's
principle in general relativity \cite{Hoy}. From the point of view
of incorporating particle physics into Einstein's theory of
gravitation, the simplest approach is to interpret the cosmological
constant $\Lambda$ in terms of quantum mechanics and the Physics of
vacuum \cite{zeld}. The $\Lambda$ term has also been interpreted in
terms of the Higgs
scalar field \cite{berg}.\\

$\Lambda$ as a function of time has also been considered by
several authors in various variable $G$ theories in different
contexts \cite{ban}. With this in view, several authors
\cite{kal,Abd} have proposed linking the variation of $G$ with
that of $\Lambda$ in the framework of general relativity. By
considering the conservation of the energy-momentum tensor of
matter and vacuum take together, many authors have invoked the
idea of a decreasing vacuum energy and hence a varying
cosmological constant $\Lambda$ with cosmic expansion in the frame
work of Einstein's theory. Present-day astronomical observations
indicate \cite{wein} that the cosmological constant $\Lambda$ is
extremely negligible being $\le 10^{-56}$ cm$^{2}$. But the value
of this constant should be $10^{50}$ times larger according to the
Glashow-Weinberg-Salam model \cite{aber} for electro-weak
unification and $10^{107}$ times larger according to GUT
\cite{Lang} for grand unification.\\

In attempt to modify the general theory of relativity, Al-Rawaf
and Taha \cite{Al} related the cosmological constant to the Ricci
scalar $R$. This is written as a built-in-cosmological constant,
i.e., $\Lambda\propto R$. Since the Ricci scalar contains a term
of the form $\frac{\ddot{a}}{a}$, one adopts this variation for
$\Lambda$ i.e., $\Lambda\propto\frac{\ddot{a}}{a}$ \cite{Arbab}.
Similarly, another two forms for $\Lambda$ have been chosen as:
$\Lambda\propto\rho$ and $\Lambda\propto\frac{\dot{a}^{2}}{a^{2}}$
\cite{Carv}; where $\rho$ is the energy density. \\

\section{The model}

The action of our model is \cite{kro}
\begin{equation}\label{1a}
S=\int d^4x{\cal L}=\int d^4x\Big\{ \sqrt{-g}\Big[
\frac{R}{G}+F(G) \Big]+{\cal L}_m \Big\},
\end{equation}
where $G$ is the Newton's gravitational constant and $F(G)$ is an
arbitrary function of $G$ while ${\cal L}_m$ is a matter
Lagrangian. From the Euler-Lagrange equation
\begin{equation}\label{1b}
\frac{\partial {\cal L}}{\partial G}=\nabla_\mu\frac{\partial
{\cal L}}{\partial(\partial_\mu G)},
\end{equation}
to obtain
\begin{equation}\label{1c}
\frac{\partial F}{\partial G}=\frac{R}{G^2}.
\end{equation}
Moreover from the variation with respect to $g_{\mu\nu}$, we have
from (\ref{1a}),
\begin{equation}\label{1d}
R_{\mu\nu}-\frac{1}{2}g_{\mu\nu}R=8\pi G
T_{\mu\nu}+g_{\mu\nu}\Big(\frac{1}{2}GF(G)\Big).
\end{equation}
Writing
\begin{equation}\label{1e}
\frac{1}{2}GF(G)=\Lambda(t),
\end{equation}
we get from (\ref{1d})
\begin{equation}\label{1f}
R_{\mu\nu}-\frac{1}{2}g_{\mu\nu}R=8\pi G
T_{\mu\nu}+g_{\mu\nu}\Lambda(t).
\end{equation}
As a background geometry, we consider a spatially homogeneous and
isotropic FRW line element
\begin{equation}\label{1}
ds^2=dt^2-a^{2}(t)\Big[\frac{dr^2}{1-kr^2}+r^2(d\theta^2+\sin^2\theta
d\phi^2) \Big],
\end{equation}
where $a(t)$ is the scale factor and $k=-1,0,+1$ is the curvature
parameter for spatially open, flat and closed universes,
respectively.

We assume that the cosmic matter is represented by the
energy-momentum tensor of a perfect fluid
\begin{equation}\label{2}
T_{\mu\nu}=(\rho+p)u_\mu u_\nu-pg_{\mu\nu}.
\end{equation}
Here $\rho$ and $p$ are the energy density and pressure of the
cosmic matter, $u_\mu$ is the four velocity satisfying $u_\mu
u^\mu=1$.

We take the equation of state
\begin{equation}\label{3}
p=(w-1)\rho,\ \ 1\leq w\leq2.
\end{equation}
Now the for the metric (\ref{1}) and stress-energy tensor (\ref{2}),
the field equations (\ref{1f}) take the form
\begin{equation}\label{4}
-2\frac{\ddot a}{a}-\Big(\frac{\dot
a}{a}\Big)^2-\frac{k}{a^2}+\Lambda=8\pi G p,
\end{equation}
\begin{equation}\label{5}
3\Big(\frac{\dot a}{a}\Big)^2+3\frac{k}{a^2}-\Lambda=8\pi G \rho.
\end{equation}
In view of vanishing of divergence of Einstein tensor, we have
\begin{equation}\label{6}
8\pi G\Big( \dot\rho+3(\rho+p)\frac{\dot a}{a} \Big)+8\pi\rho\dot
G+\dot\Lambda=0.
\end{equation}
The usual energy conservation equation $T^{\mu\nu}_{;\nu}=0$ yields
\begin{equation}\label{7}
\dot\rho+3(\rho+p)\frac{\dot a}{a}=0.
\end{equation}
Using (\ref{7}) in (\ref{6}) gives
\begin{equation}\label{8}
8\pi\rho\dot G+\dot\Lambda=0.
\end{equation}
Note that (\ref{8}) implies that $G$ is constant whenever $\Lambda$
is constant and vice-versa. Making use of (\ref{3}) in (\ref{8}), we
obtain
\begin{equation}\label{9}
\rho=\frac{C_1}{a^{3w}},
\end{equation}
where $C_1$ is a constant of integration. We can determine the value
of $C_1$ by assuming $w=w_0$, $\rho=\rho_c$ due to spatial flatness
($\rho_c$ being the critical density) at $t=t_0$, where subscript
`0' corresponds to the present value. It turns out $C_1=\rho_c
a_0^{3w_0}$. Thus (\ref{9}) becomes
\begin{equation}\label{10}
\rho=\rho_c\frac{ a_0^{3w_0}}{a^{3w}},
\end{equation}
For spatially flat case $k=0$, (\ref{4}) and (\ref{5}) become
\begin{equation}\label{11}
8\pi G p=H^2(2q-1)+\Lambda,
\end{equation}
\begin{equation}\label{12}
8\pi G \rho=3H^2-\Lambda,
\end{equation}
where $H=\frac{\dot a}{a}$ is the Hubble parameter, $q=-1-\frac{\dot
H}{H^2}$ is the deceleration parameter and $\Theta=3H$ is the
expansion scalar.

Sahni et al \cite{sahni} introduced a pair of cosmological
diagnostic pair $\{r,s\}$ which they termed as Statefinder. The two
parameters are dimensionless and are geometrical since they are
derived from the cosmic scale factor alone, though one can rewrite
them in terms of the parameters of dark energy and matter.
Additionally, the pair gives information about dark energy in a
model independent way i.e. it categorizes dark energy in the context
of background geometry only which is not dependent on the theory of
gravity. Hence geometrical variables are universal. Also this pair
generalizes the well-known geometrical parameters like the Hubble
parameter and the deceleration parameter. This pair is algebraically
related to the equation of state of dark energy and its first time
derivative.

The statefinder parameters were introduced to characterize primarily
flat universe ($k=0$) models with cold dark matter (dust) and dark
energy. They were defined as
\begin{equation}\label{111}
r\equiv\frac{\dddot a}{aH^3},
\end{equation}
\begin{equation}\label{222}
s\equiv \frac{r-1}{3(q-\frac{1}{2})}.
\end{equation}

For cosmological constant with a fixed equation of state ($w=-1$)
and a fixed Newton's gravitational constant, we have $\{1,0\}$.
Moreover $\{1,1\}$ represents the standard cold dark matter model
containing no radiation while Einstein static universe corresponds
to $\{\infty,-\infty\}$ \cite{debnath23}. In literature, the
diagnostic pair is analyzed for various dark energy candidates
including holographic dark energy \cite{zhang}, agegraphic dark
energy \cite{wei}, quintessence \cite{zhang1}, dilaton dark energy
\cite{dilaton}, Yang-Mills dark energy \cite{yang}, viscous dark
energy \cite{vis}, interacting dark energy \cite{pavon}, tachyon
\cite{shao}, modified Chaplygin gas \cite{debnath1} and $f(R)$
gravity \cite{song} to name a few.

\section{Solutions of dynamical equations}

We here assume the ansatz \cite{sing} for the variable cosmological
constant
\begin{equation}\label{13}
\Lambda(t)=\epsilon H^2,
\end{equation}
where $\epsilon$ is a constant. Making use of
(\ref{3}),(\ref{11}),(\ref{12}) and (\ref{13}), we obtain a
differential equation
\begin{equation}\label{14}
2\dot H+(3-\epsilon)wH^2=0.
\end{equation}
Solving (\ref{14}), we get
\begin{equation}\label{15}
H(t)=\frac{-2}{(\epsilon-3)wt+2C_2},
\end{equation}
where $C_2$ is an integration of constant. Note that $C_2$ can be
determined like before to get
\begin{equation}\label{16}
H(t)=\frac{2H_0}{H_0(3-\epsilon)(wt-w_0t_0)+2}.
\end{equation}
Solving for $t$ we get
\begin{equation}\label{17}
t=\frac{2}{w(\epsilon-3)H_0}\Big( 1-\frac{H_0}{H}
\Big)+t_0\frac{w_0}{w}.
\end{equation}
From (\ref{12}) and (\ref{13}), we get
\begin{equation}\label{18}
\rho=\frac{1}{8\pi G}(3-\epsilon)H^2.
\end{equation}
Integrating (\ref{16}), we get
\begin{equation}\label{19}
a(t)=C_3[(3-\epsilon)(tw-t_0w_0)H_0+2]^{\frac{2}{w(3-\epsilon)}}.
\end{equation}
Using (\ref{19}) in (\ref{10}), we get
\begin{equation}\label{20}
\rho(t)=\rho_ca_0^{3w_0}C_3^{-3w}[(3-\epsilon)(tw-t_0w_0)H_0+2]^{\frac{6}{(\epsilon-3)}}.
\end{equation}
Using (\ref{19}) in (\ref{13}), we get
\begin{equation}\label{21}
\Lambda(t)=\frac{4\epsilon
H_0^2}{[H_0(3-\epsilon)(tw-t_0w_0)+2]^2}.
\end{equation}
Making use of (\ref{21}) and (\ref{20}) in (\ref{8}) to get
\begin{equation}\label{22}
G(t)=\frac{H_0^2(3-\epsilon)[(3-\epsilon)(tw-t_0w_0)H_0+2]^{\frac{2\epsilon}{3-\epsilon}}}
{2\pi \rho_ca_0^{3w_0}C_3^{-3w}}.
\end{equation}
The deceleration parameter becomes
\begin{equation}\label{23}
q=-1+\frac{w(3-\epsilon)}{2}.
\end{equation}
For an expanding universe, $q\leq-1$ which constrains $\epsilon<3$.
The expansion scalar takes the form
\begin{equation}\label{24}
\Theta=\frac{6H_0}{H_0(3-\epsilon)(wt-w_0t_0)+2}.
\end{equation}

For flat universe, the density parameter can be obtained as

\begin{equation}\label{24}
\Omega=\frac{8\pi G\rho}{3H^{2}}=1-\frac{\epsilon}{3}
\end{equation}

We now calculate the statefinder parameters $\{r,s\}$ as well
\begin{equation}\label{25}
r=\frac{\dddot a}{aH^3},\ \ \ s=\frac{r-1}{3(q-\frac{1}{2})}.
\end{equation}
The parameters take the form
\begin{equation}\label{26}
r=\frac{1}{2}[1+w(\epsilon-3)][2+w(\epsilon-3)],\ \ \
s=\frac{w}{3}(3-\epsilon).
\end{equation}
Note that for $\epsilon=3$, $\{r,s\}$=$\{1,0\}$ representing a
static cosmological constant.

\subsection{Universe containing only dust}

The matter (baryonic and non-baryonic) satisfies the EoS parameter
$w=1$. Thus cosmological parameters take the form

\begin{eqnarray}
a(t)&=&C_3[(3-\epsilon)(t-t_0w_0)H_0+2]^{\frac{2}{3-\epsilon}},\\
\rho(t)&=&\rho_ca_0^{3w_0}C_3^{-3}[(3-\epsilon)(t-t_0w_0)H_0+2]^{\frac{6}{\epsilon-3}},\\
\Lambda(t)&=&\frac{4\epsilon
H_0^2}{[H_0(3-\epsilon)(t-t_0w_0)+2]^2},\\
G(t)&=&\frac{H_0^2(3-\epsilon)[(3-\epsilon)(t-t_0w_0)H_0+2]^{\frac{2\epsilon}{3-\epsilon}}}
{2\pi \rho_ca_0^{3w_0}C_3^{-3}},\\
q&=&\frac{1-\epsilon}{2},\\
\Theta&=&\frac{6H_0}{H_0(3-\epsilon)(t-w_0t_0)+2},\\
r&=&\frac{1}{2}(\epsilon-2)(\epsilon-1),\\
s&=&\frac{1}{3}(3-\epsilon).
\end{eqnarray}
\subsection{Universe containing only radiation}

The radiation satisfies the EoS parameter $w=4/3$. Thus cosmological
parameters take the form
\begin{eqnarray}
a(t)&=&C_3\Big[(3-\epsilon)\Big(\frac{4}{3}t-t_0w_0\Big)H_0+2\Big]^{\frac{3}{2(3-\epsilon)}},\\
\rho(t)&=&\rho_ca_0^{3w_0}C_3^{-4}\Big[(3-\epsilon)\Big(\frac{4}{3}t-t_0w_0\Big)H_0+2\Big]^{\frac{6}{(\epsilon-3)}},\\
\Lambda(t)&=&\frac{4\epsilon
H_0^2}{\Big[(3-\epsilon)\Big(\frac{4}{3}t-t_0w_0\Big)H_0+2\Big]^2},\\
G(t)&=&\frac{H_0^2(3-\epsilon)\Big[(3-\epsilon)\Big(\frac{4}{3}t-t_0w_0\Big)H_0+2\Big]^{\frac{2\epsilon}{3-\epsilon}}}
{2\pi \rho_ca_0^{3w_0}C_3^{-4}},
\end{eqnarray}
\begin{eqnarray}
q&=&1-\frac{2\epsilon}{3},\\
\Theta&=&\frac{6H_0}{(3-\epsilon)\Big(\frac{4}{3}t-t_0w_0\Big)H_0+2},\\
r&=&\frac{1}{9}(4\epsilon-9)(2\epsilon-3),\\
s&=&\frac{4}{9}(3-\epsilon).
\end{eqnarray}

\subsection{Universe containing only stiff matter}

The stiff fluid satisfies the EoS parameter $w=2$. Thus cosmological
parameters take the form
\begin{eqnarray}
a(t)&=&C_3[(3-\epsilon)(2t-t_0w_0)H_0+2]^{\frac{1}{3-\epsilon}},\\
\rho(t)&=&\rho_ca_0^{3w_0}C_3^{-6}[(3-\epsilon)(2t-t_0w_0)H_0+2]^{\frac{6}{\epsilon-3}},\\
\Lambda(t)&=&\frac{4\epsilon
H_0^2}{[H_0(3-\epsilon)(2t-t_0w_0)+2]^2},\\
G(t)&=&\frac{H_0^2(3-\epsilon)[(3-\epsilon)(2t-t_0w_0)H_0+2]
^{\frac{2\epsilon}{3-\epsilon}}} {2\pi \rho_ca_0^{3w_0}C_3^{-6}},\\
q&=&2-\epsilon,\\
\Theta&=&\frac{6H_0}{H_0(3-\epsilon)(2t-w_0t_0)+2},\\
r&=&(2\epsilon-5)(\epsilon-2),\\ s&=&\frac{2}{3}(3-\epsilon).
\end{eqnarray}

\section{Some consequences}

In this section, we'll discuss lookback time, proper distance,
luminosity distance and angular diameter as the following
\cite{arb}.

\subsection{Lookback time}

If a photon emitted by a source at the instant $t$ and received at
the time $t_{0}$ then the photon travel time or the lookback time
$t-t_{0}$ is defined by
\begin{equation}
t-t_{0}=\int_{a_{0}}^{a}\frac{da}{\dot{a}},
\end{equation}
where $a_{0}$ is the present value of the scale factor of the
universe and can be obtained from (\ref{19}) as (at $t=t_{0}$,
$w=w_{0}$)
\begin{equation}
a_{0}=2^{\frac{2}{w_{0}(3-\epsilon)}}~C_{3}.
\end{equation}
The redshift $z$ can be defined by
\begin{equation}
1+z=\frac{a_{0}}{a},
\end{equation}
which simplifies to
\begin{equation}
t-t_{0}=\frac{2}{w(\epsilon-3)H_{0}}+\left(\frac{w_{0}}{w}-1\right)t_{0}
+\frac{2^{^{\frac{w_{0}}{w}}}}{w(3-\epsilon)H_{0}}(1+z)^{-\frac{w(3-\epsilon)}{2}}.
\end{equation}
Since, $w>0$ and $\epsilon<3$ always, in our assumption. So for
early universe i.e., for $z\rightarrow \infty$ we get
\begin{equation}
t-t_{0}\approx
\frac{2}{w(\epsilon-3)H_{0}}+\left(\frac{w_{0}}{w}-1\right)t_{0},
\end{equation}
and for late universe i.e., for $z\rightarrow -1$ we get
\begin{equation}
t-t_{0}\approx
\frac{2^{^{\frac{w_{0}}{w}}}}{w(3-\epsilon)H_{0}}(1+z)^{-\frac{w(3-\epsilon)}{2}}.
\end{equation}

\subsection{Proper distance}

If a photon emitted by a source and received by an observer at time
$t_{0}$ then the proper distance between them is defined by
\begin{equation}
d=a_{0}\int_{a}^{a_{0}}\frac{da}{a\dot{a}}=a_{0}\int_{t}^{t_{0}}\frac{dt}{a},
\end{equation}
which simplifies to
\begin{equation}
d=\frac{2^{^{\frac{w}{w_{0}}}}(1+z)^{1+\frac{w(\epsilon-3)}{2}}}
{(2+w(\epsilon-3))H_{0}}-\frac{2}{(2+w_{0}(\epsilon-3))H_{0}}.
\end{equation}

\subsection{Luminosity distance}

If $L$ be the total energy emitted by the source per unit time and
$\ell$ be the apparent luminosity of the object then the luminosity
distance is defined by
\begin{equation}
d_{L}=\left(\frac{L}{4\pi\ell}\right)^{\frac{1}{2}}.
\end{equation}
Now for our model,
\begin{equation}
d_{L}=d(1+z)=\frac{2^{^{\frac{w}{w_{0}}}}(1+z)^{2+\frac{w(\epsilon-3)}{2}}
}{(2+w(\epsilon-3))H_{0}}-\frac{2(1+z)}{(2+w_{0}(\epsilon-3))H_{0}}.
\end{equation}

\subsection{Angular diameter}

The angular diameter of a light source of proper distance $D$
observed at $t_{0}$ is defined by
\begin{equation}
\delta=\frac{D(1+z)^{2}}{d_{L}}.
\end{equation}
The angular diameter distance ($d_{A}$) is defined as the ratio of
the source diameter to its angular diameter as
\begin{equation}
d_{A}=\frac{D}{\delta}=d_{L}(1+z)^{-2},
\end{equation}
which is simplifies to
\begin{equation}
d_{A}=\frac{2^{^{\frac{w}{w_{0}}}}(1+z)^{\frac{w(\epsilon-3)}{2}}
}{(2+w(\epsilon-3))H_{0}}-\frac{2(1+z)^{-1}}{(2+w_{0}(\epsilon-3))H_{0}}.
\end{equation}
The angular diameter distance is maximum at
\begin{equation}
z_{max}=\left(\frac{2^{^{2-\frac{w}{w_{0}}}}}{w(3-\epsilon)}~~
\frac{2+w(\epsilon-3)}{2+w_{0}(\epsilon-3)}\right)^{\frac{2}{2+w(\epsilon-3)}}
  -1,
\end{equation}
and hence the maximum angular diameter distance will be
\begin{equation}
\left(d_{A}\right)_{max}=\frac{2}{H_{0}}~\left(2^{^{2-\frac{w}{w_{0}}}}
\right)^{-\frac{2}{2+w(\epsilon-3)}}~\left(\frac{2+w(\epsilon-3)}{w(3-\epsilon)
(2+w_{0}(\epsilon-3))}\right)^{\frac{w(\epsilon-3)}{2+w(\epsilon-3)}}.
\end{equation}

\section{Conclusions}

In this work, we have considered a cosmological model of the
homogeneous and isotropic FRW universe with variable Newton's
gravitational constant and cosmological constant. The solutions
have been obtained for flat model with barotropic fluid and some
particular form of cosmological constant (i.e., $\Lambda=\epsilon
H^{2}$). The cosmological parameters and deceleration parameter
have been obtained for dust, radiation and stiff perfect fluid.
The statefinder parameters are analyzed for these three types of
fluids and have shown that these depends only on $w$ and
$\epsilon$. From figure 1, we see that $q$ decreases with $w$
decreases and $\epsilon$ increases. Also figures 1-3 describe the
natures of statefinder parameters for these three types of fluids.
Further the lookback time, proper distance, luminosity distance
and angular diameter distance have also been investigated.
The maximum angular diameter distance has also been found in our model.\\

\pagebreak\newpage
\begin{figure}
\includegraphics[scale=.9]{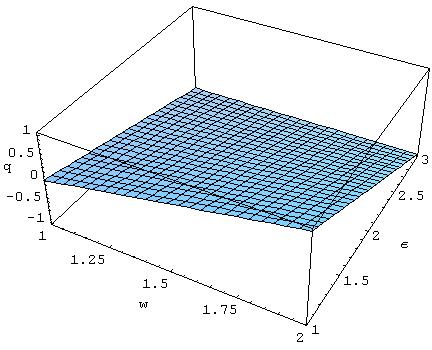}\\
\caption{The deceleration parameter $q$ is plotted for different
choices of $\epsilon$ and $w$.}
\end{figure}

\begin{figure}
\includegraphics[scale=.9]{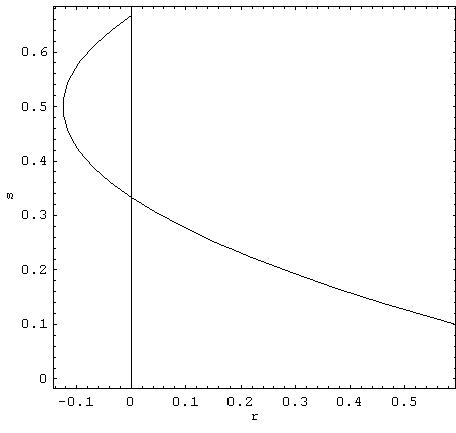}\\
\caption{The statefinder parameters are plotted for $w=1$ and
$\epsilon=1...3$.}
\end{figure}

\begin{figure}
\includegraphics[scale=.9]{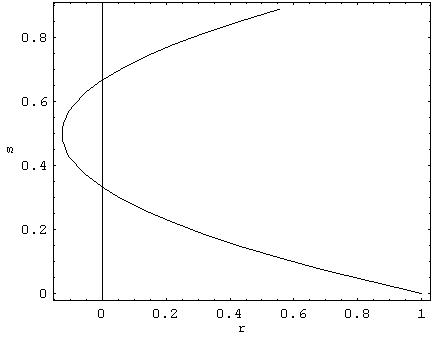}\\
\caption{The statefinder parameters are plotted for $w=4/3$ and
$\epsilon=1...3$.}
\end{figure}

\begin{figure}
\includegraphics[scale=.9]{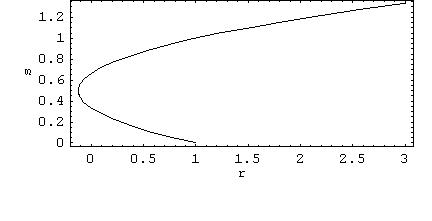}\\
\caption{The statefinder parameters are plotted for $w=2$ and
$\epsilon=1...3$.}
\end{figure}


\begin{thebibliography}{99}

\bibitem{wess} P. S. Wesson (1978), {\it Cosmology and Geophysics} (Oxford :
Oxford University Press ); P. S. Wesson (1980), {\it Gravity,
Particles and Astrophysics} ( Dordrecht : Rieded ).

\bibitem{tiwari} R. K. Tiwari, \textit{Astrophys. Space Sci.} \textbf{321} 147 (2009);
 M. Jamil et al, \textit{Eur. Phys. J. C} \textbf{60} 149 (2009); M.
 Jamil et al, \textit{Phys. Lett. B} \textbf{679} 172 (2009); M. R.
 Setare and M. Jamil, \textit{JCAP} \textbf{1002} 010 (2010).

\bibitem{Dir} P. A. M. Dirac, {\it Proc. R. Soc. A} {\bf 165} 119 (1938) ;
{\bf 365} 19 (1979) ; {\bf 333} 403 (1973); The General Theory of
Relativity (New York : Wiley) 1975.

\bibitem{Hoy} F. Hoyle and J. V. Narlikar, {\it Proc. R. Soc. A} {\bf 282} 191
(1964); {\it Nature} {\bf 233} 41 (1971); C. Brans and R. H.
Dicke, {\it Phys. Rev.} {\bf 124} 925 (1961).

\bibitem{zeld} Y. B. Zeldovich, {\it Sov. Phys.- JETP} {\bf 14} 1143 (1968);
{\it Usp. Fiz. Nauk} {\bf 11} 384 (1968); P. J. E. Peebles and B.
Ratra, {\it Astrophys. J.} {\bf 325} L17 (1988).

\bibitem{berg} P. G. Bergmann, {\it Int. J. Theor. Phys.} {\bf 1} 25
(1968); R. V. Agoner, {\it Phys. Rev. D} {\bf 1} 3209 (1970).

\bibitem{ban} A. Banerjee, S. B. Dutta Chaudhuri and N. Banerjee, {\it Phys.
Rev. D} {\bf 32} 3096 (1985); O. Bertolami {\it Nuovo Cimento}
{\bf 93B} 36 (1986); {\it Fortschr. Phys.} {\bf 34} 829 (1986);
Abdussattar and R. G. Vishwakarma, {\it Class. Quantum Grav.} {\bf
14} 945 (1997).

\bibitem{kal} D. Kaligas, P. Wesson and C. W. F. Everitt, {\it Gen. Rel.
Grav.} {\bf 24} 351 (1992).

\bibitem{Abd} A - M. M. Abdel Rahaman, {\it Gen. Rel. Grav.} {\bf 22} 655
(1990); M. S. Berman, {\it Gen. Rel. Grav.} {\bf 23} 465 (1991);
A. Beesham, {\it Int. J. Theor. Phys.} {\bf 25} 1295 (1986).

\bibitem{wein} Weinberg, S. (1972). Gravitation and Cosmology, (Wiley, New
York).

\bibitem{aber} Abers, E. S., and Lee, B. W. (1973). Phys. Rep. 9, 1.

\bibitem{Lang} Langacker, P. (1981). Phys. Rep. 72, 185.

\bibitem{Al} A. S. Al-Rawaf and M. O. Taha, {\it Gen. Rel. Grav.} {\bf
28} 935 (1996).

\bibitem{Arbab} Arbab I. Arbab, {\it Class. Quantum Grav.} {\bf 20} 93
(2003); {\it Astrophys. Space Sci.} {\bf 291} 141 (2004).

\bibitem{Carv} J. C. Carvalho, J. A. S. Lima and L. Waga, {\it Phys. Rev. D}
{\bf 46} 2404 (1992).

\bibitem{kro} K.D. Krori et al, Gen. Relativ. Grav. 32 (2000) 1439.

\bibitem{sahni} V. Sahni et al, JETP Lett. 77 (2003) 201.

\bibitem{debnath23} U. Debnath, Class. Quant. Grav. 25 (2008) 205019.

\bibitem{zhang} X. Zhang, Int. J. Mod. Phys. D 14 (2005) 1597.

\bibitem{wei} H. Wei and R.G. Cai, Phys. Lett. B 655 (2007) 1.

\bibitem{zhang1} X. Zhang, Phys. Lett. B 611 (2005) 1.

\bibitem{dilaton} J.Z. Huang et al,  Astrophys. Space Sci. 315 (2008)
175.
\bibitem{yang} W. Zhao, Int. J. Mod. Phys. D 17 (2008) 1245.

\bibitem{vis} M. Hu and X.H. Meng, Phys. Lett. B 635 (2006) 186.

\bibitem{pavon} W. Zimdahl and D. Pavon, Gen. Rel. Grav. 36 (2004)
1483.

\bibitem{shao} Y. Shao and Y. Gui, Mod. Phys. Lett. A 23 (2008) 65.

\bibitem{debnath1} W. Chakraborty and U. Debnath, Mod. Phys. Lett. A 22 (2007)
1805.

\bibitem{song} S. Li et al, arXiv:1002.3867 [astro-ph.CO]

\bibitem{sing} T. Singh et al, Gen. Relativ. Grav. 30 (1998) 573.

\bibitem{arb} A. I. Arbab, astro-ph/9810239.




\end{thebibliography}
\end{document}